\documentclass{article}
\usepackage{spconf,graphicx}
\usepackage{amssymb}
\usepackage{amsmath}
\usepackage{psfrag}

\newcommand{\B}[1]{\boldsymbol{#1}}

\title{Sparse Linear Precoders for mitigating 
nonlinearities in massive MIMO}
\name{Amine Mezghani$^1$, Daniel Plabst$^4$, Lee A. Swindlehurst$^2$, Inbar Fijalkow$^3$, Josef A. Nossek$^4$ } 
\address{$^1$University of Manitoba, MB, Canada; ~~~~~~
   $^3$ETIS, CY Cergy Paris Univ, ENSEA, CNRS, Paris, France \\
		$^2$University of California, Irvine, CA, USA; ~~~~~~~~~~~~~~~
	$^4$Technische Universit\"at M\"unchen, Germany}
\begin{document}
%
\maketitle
\begin{abstract}

Dealing with nonlinear effects of the radio-frequency (RF) chain is a key issue in the realization of very large-scale multi-antenna  (MIMO) systems. Achieving the remarkable gains possible with massive MIMO 
requires that the signal processing algorithms systematically take into account these effects. Here,  we present a computationally-efficient linear precoding method satisfying the requirements for low peak-to-average power ratio (PAPR) and low-resolution D/A-converters (DACs). The method is based on a sparse regularization of the precoding matrix and offers advantages in terms of precoded signal PAPR as well as processing complexity. Through simulation, we find that the method substantially improves conventional linear precoders. 

\end{abstract}
%
%
\section{Introduction}
\label{sec:intro}

The use of large-scale multiple antenna systems,  known as Massive MIMO, has emerged as a leading concept for advancing wireless communications infrastructure. The design of such systems, however, comes with inherent challenges in terms of analog hardware and computational complexity. The rapidly growing pace of scale in antenna systems for communications necessitates methods to reduce the RF hardware complexity and power consumption, such as reducing the resolution of the data converters and operating the power-amplifiers with constant-envelope signals. In particular, the power amplifier (PA) typically accounts for a substantial amount of the power consumption in a base station (BS) \cite{Blume2010}.  Therefore, it is desirable in terms of power efficiency to run the PA in the saturation region where most of the available PA power is radiated and less power is lost as heat.


Operating PAs in the saturation region and using low-resolution D/A-converters (DACs), however, implies high distortions and nonlinearities that are introduced to the signals, calling for new processing techniques to mitigate their effects.
In \cite{mollen_2016}, the impact of power amplifier nonlinearity has been analyzed and it was shown that both single-carrier (SC) and OFDM transmission are affected similarly if conventional linear precoding is used since the peak-to-average power ratio (PAPR) increases with multi-user processing \cite{Jedda2019}.  Standard and improved linear zero-forcing and minimum-mean-squared error (MMSE) precoder designs for (massive) MIMO with low-resolution DACs and their performance have been studied in  \cite{Mezghani_2009, Usman_2016,Saxena_2016_2,DeCandido2019,Li2017,amodh_2020}, showing satisfactory performance for small loading factors and well-conditioned channels such as i.i.d. channels. Achievable rates for multi-user MISO systems with low-resolution D/A converters were considered in \cite{Kakkavas_2016}. Nonlinear \emph{Tomlinson-Harashima Precoding} has been considered in \cite{Mezghani_2008_G} for low-resolution DACs, and achieves better performance than pure linear methods. Other works such as  \cite{Jedda_2016,Jacobsson_2016,Jacobsson_2016_2,Swindlehurst_2017,Studer2013,Shao_2019,Lopes2020} show that nonlinear symbol-by-symbol precoding schemes outperform linear precoders under hardware impairments at the cost of an increased computational complexity.
 
Motivated by the simplicity of linear methods, we reconsider the problem of optimizing linear precoding techniques and present a new approach where SC transmission can still preserve its advantage in terms of PAPR compared to OFDM despite the multi-user processing. The new precoders are used in particular for mitigating the effects of constant envelope transmission as well as 1-bit DACs. To this end, a new cost functions with $\ell_1$ regularization is investigated. The new class of linear precoders based on $\ell_1$ matrix norm optimization significantly outperforms conventional designs for QPSK as well as higher-order modulations and allows for reduced computational complexity.

\section{System Model}
\label{sec:format}

Consider a downlink system with nonlinear transmitters:  
\begin{equation}
  \hat{s}=\B{H} \cdot \mathcal{Q} (\B{P}\cdot \B{s}) +\B{\eta},
\end{equation}
where $\B{H} \in \mathbb{C}^{K\times N}$ is the channel matrix with $K$ single-antenna users and $N$ transmit antenna, while  $\B{P} \in \mathbb{C}^{N\times K}$ is the linear precoder and $\mathcal{Q}(\cdot)$ represent the transmit nonlinearity. The transmit symbol vector $\B{s}$ is assumed to be either an OFDM or a single-carrier signal and the noise vector $\B{\eta}$ is i.i.d. circular symmetric complex Gaussian with variance $\sigma_\eta^2$.

One possible design criterion for $\B{P}$ is that it be zero-forcing, i.e, $\B{H}\cdot \B{P}= {\bf I}$ if quantization is ignored. However since usually $N>K$, the solution is not unique. Thus the question that arises is what is an appropriate solution  for mitigating nonlinearities such us clipping at the power amplifier or one-bit quantization at the DAC. We will focus  on two types of nonlinearities: (i) One-bit DAC quantization:
\begin{equation}
\mathcal{Q}(\B{y})=\frac{1}{\sqrt{2}}{\rm sign}({\rm Re} \{{\B{y}}\})+\frac{\rm j}{\sqrt{2}}{\rm sign}({\rm Im} \{{\B{y}}\}).
\end{equation}
and (ii) constant envelope transmission 
\begin{equation}
\mathcal{Q}(\B{y})=\exp\left( {\rm j} \angle \B{y} \right),
\end{equation}
where the nonlinear operation is applied element-wise. In both cases, we obtain constant instantaneous power per antenna equal to one.  We study in the next section the impact of OFDM or single-carrier processing on the PAPR and how standard linear multiuser precoding techniques affect those signals.  

\section{PAPR Comparison between single-carrier and OFDM systems}
\label{sec:pagestyle}

In this section, we show that for the same power spectral density and the same performance under AWGN (with perfect orthogonal transmission), the worst-case peak power in the single-carrier (SC) systems grows only logarithmically with the spectral rolloff (steepness of the spectrum in the transmission band), whereas the worst-case peak power in OFDM grows linearly with the rolloff (i.e., proportional to the number of sub-carriers).

\subsection{OFDM vs. block-wise single-carrier transmission}
We compare OFDM and block-wise single-carrier modulation (also known as SC-FDMA in LTE) assuming an oversampling (interpolation) factor of two. In OFDM, a block of $M$ data symbols (consisting in general of QAM symbols) is mapped to the lower frequency points of a $2M$-point IFFT with zero-padding. In the block-wise  single-carrier scheme, an additional $M$-point FFT is used to precode the data, resulting in a kind of data mapping in the time domain. The single-carrier processing at the transmitter can also be regarded as a cyclic convolution of data blocks with the impulse response of a discrete rectangular filter with a window size of $M$ out of $2M$ frequency points,
\begin{equation}
  P_k=\left\{ 
\begin{array}{ll}
	\frac{1}{\sqrt{M}} & \textrm{~for~} k < \frac{M}{4}  \textrm{~or~} 2M-k < \frac{M}{4}-1   \\
	\frac{1}{2\sqrt{M}} & \textrm{~for~} k = \frac{M}{4}  \textrm{~or~} 2M-k = \frac{M}{4}-1, 
\end{array}
  \right.
\end{equation}
and zero elsewhere, with $0 \leq k < 2M$. The impulse response of the cyclic pulse shaping filter is given by applying an IFFT 
\begin{equation}
 p[n] =\left\{ 
\begin{array}{ll} 
1 & \textrm{~for~} n =0 \\
 \frac{\sin\left(\pi \frac{n}{2}\right)}{ M \sin\left(\pi \frac{n}{2M}\right)  } & \textrm{~otherwise} \end{array} \right. , \quad 0 \leq n < 2M.
\end{equation}
Both systems have the same spectral density since the additional $M$-point FFT in SC-FDMA is a unitary transformation that does not change the second-order statistics. The resulting steepness of the spectral confinement is proportional to $M$. Also, in the AWGN case, both systems provide perfect orthogonality between data symbols and thus achieve the same BER performance. However, we show that these systems have significantly different behavior in terms of PAPR. 

\subsection{Peak-to-average power analysis}
For simplicity, in this brief analysis we assume the input to be BPSK. For OFDM, the maximum peak-to-average-power is $M$ and is achieved when all the inputs are identical. In the single-carrier case the peak power is achieved at the  intermediate time instants between two consecutive symbols, when all symbols add coherently. Therefore, we get the peak power
 \begin{equation}
 \begin{aligned}
    \left(\sum_{n=0}^{M-1} |p[2n+1]|\right)^2 &= \left(\sum_{n=0}^{M-1} \left| \frac{\sin\left(\pi \frac{2n+1}{2}\right)}{ M \sin\left(\pi \frac{2n+1}{2M}\right)  } \right|\right)^2 \\
    &= \left(\sum_{n=0}^{M-1}  \frac{1}{ M \sin\left(\pi \frac{2n+1}{2M}\right)  } \right)^2 \\
    &\leq \left( \int\limits_{\frac{1}{4M}}^{1-\frac{1}{4M}} \frac{1}{\sin(\pi x)} {\rm d}x  \right)^{\! 2} \!\! + \! \frac{1}{M \sin (\frac{\pi}{2M}) }\\
    &\approx \frac{4}{\pi^2}\left(\log\left(\sin(\frac{\pi}{8M})\right)  \right)^2 \\
    &\approx \frac{4}{\pi^2}\left(\log\left(\frac{\pi}{8M}\right)  \right)^2,
  \end{aligned}
 \end{equation}
where the approximations hold for large $M$. Clearly, we see that the peak power of the single-carrier system only increases logarithmically with $M$, showing its superiority compared to OFDM in terms of PAPR. Standard linear multiuser precoding, however, diminishes this property due to the central limit theorem \cite{Jedda2019}. Our goal is to preserve this advantage in massive MIMO using sparse precoding.

\section{Sparse precoding design}
\label{sec:typestyle}
The most common choice for the zero-forcing precoder minimizes the Frobenius norm of $\B{P}$, i.e., the transmit power:
\begin{equation}
  \min\limits_{\B{P}} \left\| \B{P} \right\|_{2,2}^2 \quad  \textrm{s.t. }  \B{H}\cdot \B{P}= {\bf I},
\end{equation}
where the $\ell_{p,q}$ matrix norm is defined as
\begin{equation}
\left\| \B{A} \right\|_{p,q}^q=  \sum_j \left( \sum_i |a_{i,j}|^p \right)^{q/p}.
\end{equation}
The solution to this problem is famously given by the pseudo-inverse of $\B{H}$
\begin{equation}
    \B{P}= \B{H}^{\rm H} \left( \B{H} \B{H}^{\rm H} \right)^{-1},
\end{equation}
which aims at minimizing the transmit power while fulfilling the zero-forcing requirement, or equivalently it maximizes the signal-to-noise ratio at the user terminals for a given power constraint. However, this solution results in a dense matrix $\B{P}$ (even when $\B{H}$ is sparse), leading to the fact that the distribution of $\B{P}\cdot \B{s}$ converges element-wise to a Gaussian distribution with high kurtosis. This happens even when  $\B{s}$ is QPSK, resulting in high distortion at the DACs or amplifiers and eventually producing an error floor at high SNR. Therefore, an alternative formulation is presented in the next section.

\section{New approach: Formulation based on the $\ell_{1,2}$ norm}

If the input $\B{s}$ is a QPSK signal and the desired output of the precoder $\B{Ps}$ should be constant envelope or even QPSK (in the case of one-bit DACs), then we propose to design the precoder to have sparse rows, in order to preserve the statistical properties of  $\B{s}$ as much as possible after precoding. To this end, we use the $\ell_{1,2}$ matrix norm 
\begin{equation}
  \min\limits_{\B{P}} \left\| \B{P}^{\rm T} \right\|_{1,2}^2=\sum_i \Big( \sum_j |p_{i,j}| \Big)^2  \quad  \textrm{s.t. }  \B{H}\cdot \B{P}= {\bf I}.
\end{equation}
This problem is convex and is solved efficiently. For constant envelope signaling, $|p_{i,j}|$ is the Euclidean norm.
For the one-bit DAC case, as it is applied separately to the inphase and quadrature components, it is appropriate to use a real valued representation of the channel 
and to redefine the norm as
\begin{equation}
|p_{i,j}| \stackrel{\rm 1-bit}{=}   |{\rm Re}\{p_{i,j}\}| + |{\rm Im}\{p_{i,j}\}| . 
\end{equation}
Another important advantage of sparse precoding is the reduced computation for the matrix-vector multiplication. 

\subsection{Combined $\ell_{1,2}/\ell_{2,2}$ regularization}
\begin{figure}
  \psfrag{-10*log10}[c][c]{\tiny $-10\log_{10} \sigma_\eta^2$}
\psfrag{BER}[c][c]{\tiny BER}
  \centerline{\includegraphics[width=5.5cm]{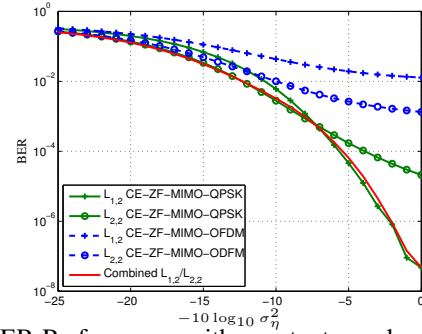}}
  \vspace{-0.5cm}
\caption{BER Performance with constant envelope precoding for the downlink with $K = 5$ users and $N = 30$ antennas, QPSK.}
\label{fig_2}
\end{figure}

As we will see in the simulations, the  $\ell_{2,2}$ formulation performs better at low SNR while the $\ell_{1,2}$ approach performs better at high SNR. This is because the quantization effects dominate the noise only for high SNR values. This suggests the combination of both norms in a way that the $\ell_{2,2}$ norm  is effective at low SNR while the $\ell_{1,2}$ norm becomes active at high SNR values, similar to the \emph{elastic net} approach of \cite{Zou2005}:
\begin{equation}
  \min\limits_{\B{P}}   \left\| \B{H}\B{P} -{\bf I}  \right\|_{2,2}^2 + \frac{1}{2} \left( \frac{K\cdot \sigma_\eta^2}{N}  \left\| \B{P}^{\rm T} \right\|_{2,2}^2 +\lambda\left\| \B{P}^{\rm T} \right\|_{1,2}^2    \right),
\end{equation}
where $\lambda$ is a constant corresponding to the value $K\cdot \sigma_{\eta,\rm cross}^2/N$ at exactly the crossing point of both BER curves (c.f. Fig.~\ref{fig_2}).
This combined $\ell_{1,2}/\ell_{2,2}$ regularized optimization can be solved iteratively using the iterative shrinkage-thresholding algorithm (ISTA) \cite{Yang_2010} with an appropriate stepsize $\mu$ 
\begin{equation}
\begin{aligned}
 &\B{P}^{\ell+1}= {\rm exp}\left({\rm j}\angle(\B{P}^\ell - \mu \B{\Delta}^\ell)\right) \circ \\
 &~~~~~~ \max \Big( \left| \B{P}^\ell -\mu \B{\Delta}^\ell \right|-\mu \frac{\lambda}{2} \cdot (\sum_k |\B{p}_{k}^{\ell}|)\cdot \B{1}^{\rm T}, \B{0} \Big),
 \end{aligned}
 \label{grad_up}
\end{equation}
where $\circ$ represents the Hadamard product and 
\begin{equation}
\B{\Delta}^\ell= \left( \B{H}^{\rm H} \B{H} +  \frac{1}{2}  \frac{K\cdot \sigma_\eta^2}{N} \cdot    {\bf I} \right) \B{P}^\ell  -\B{H}^{\rm H}.
\end{equation}
 In the iterative formula (\ref{grad_up}), the absolute values are taken elementwise (note that for 1-bit we define $|x+y{\rm j}|=|x|+|y|$) and $\B{p}_{k}$ are the column vectors of $\B{P}$. 
\subsection{Superposition coding for higher-order modulation}

Using higher-order modulation such as 16-QAM for better spectral efficiency while the base station only employs 1-bit DACs at each antenna may seem to be contradictory at first glance. However since the number of antennas is much larger than the number of users, such an idea can be achieved in principle with an appropriate transmit strategy. We propose such a method, still based on linear precoding, which is applicable for square QAM constellations. To this end, we apply the concept of superposition coding \cite{DeCandido2019}, to reformulate the problem again based on QPSK  symbols that superimpose "over-the-air" through the channel $\B{H}$ to form the desired QAM constellation at the user terminals. For simplicity, let us consider the 16-QAM case, which can be represented as the sum of two QPSK signals: One least significant symbol (LSB) and one most  significant symbol (MSB)
\begin{equation}
\B{s}_{\rm 16QAM}= (  {\bf I}_K \otimes [1,~2]) \cdot \B{s}_{\rm QPSK} = \B{\Pi}\cdot \B{s}_{\rm QPSK},
\end{equation}
where $\B{s}_{\rm QPSK} \in \{\pm 1\pm {\rm j} \}^{2K}$. Now, instead of designing a precoder matrix $\B{P}$ to be applied directly to $\B{s}_{\rm 16QAM}$, we design a larger precoder $\B{P} \in \mathbb{C}^{N\times 2K}$ that is applied to $\B{s}_{\rm QPSK}$, and we let the corresponding signals superimpose through the channel and form the desired signal $\B{s}_{\rm 16QAM}$ at the user terminals. By doing so and additionally encouraging $\B{P}$ to be sparse, we can produce a binary-like signal at the output of the precoder and mitigate the issue of the 1-bit DACs, while at the receiver side a higher-order modulated signal is generated through the channel.  

Unfortunately, with the new constraint $\B{H}\B{P}=\B{\Pi}$, the Restricted Isometry Property (RIP) condition is not valid for this problem meaning that the $\ell_1$ solution might not be sparse as desired. In fact, due to symmetry considerations (same channel for both MSB and LSB bits), the solution will present the same symmetry, i.e., the precoding vector for the most significant bit is just twice the one for the LSB bit. To break this symmetry, we propose to add some small non-convex perturbation of the original problem in order to encourage sparse solutions having different precoding vectors for both bits:
\begin{equation}
\begin{aligned}
  &\min\limits_{\B{P}}   \left\| \B{H}\B{P} -\B{\Pi}  \right\|_{2,2}^2 + \\
  &\frac{1}{2} \left( \frac{K\cdot \sigma_\eta^2}{N}  \left\| \B{P}^{\rm T} \right\|_{2,2}^2 +\lambda \left(\left\| \B{P}^{\rm T} \right\|_{1,2}^2   + {\rm tr} (\B{P}{\bf J}\B{P}^{\rm H}) \right) \right),
\end{aligned}
\end{equation}
with  $ {\bf J}= {\bf I}_K  \otimes ({\bf 1} {\bf 1}^{\rm T}-{\bf I}_2)$. The term $ {\rm tr} (\B{P}{\bf J}\B{P}^{\rm H}) $ represents the non-convex perturbation for breaking the symmetry. Again, the iterative shrinkage-thresholding algorithm can be used to solve this optimization (locally).


\section{Simulation Results}
\label{sec:foot}
\begin{figure}
  \centering
  \psfrag{eta}[c][c]{$\eta$}
\psfrag{P}[c][c]{$P({\rm output~power} > \eta)$}
  \centerline{\includegraphics[width=6.4cm]{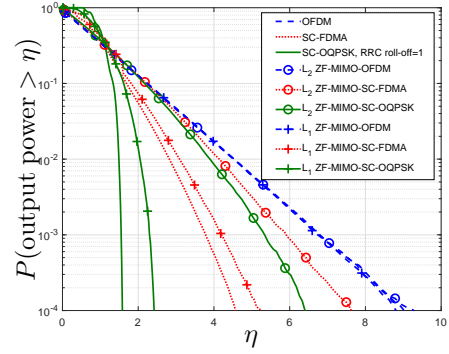}}
\caption{OFDM vs. single-carrier in terms of the CCDF of instantaneous output power for QPSK input and $M = 128$, $N=30$ antennas, and $K=5$ users.}
\label{fig_1}
\end{figure}

We consider first a system with $N=30$ antennas,  $K=5$ users, and QPSK constellations.  The channel entries are $h_{ij} \sim \mathcal{CN}(0,1)$. In the simulations, we used a conservative value for the stepsize $\mu=0.01$.  Accelerated ISTA versions (such as FISTA) can be also used for significantly faster convergence rate \cite{Yang_2010}. In Fig.~\ref{fig_1}, the complementary cumulative distribution function (CCDF) of the output power is shown for the new precoding technique with $\ell_1$ regularization compared to the conventional ZF-technique for different modulation formats. For comparison, the CCDF without precoding is also shown. With the new precoder, the PAPR is only increased marginally, while SC-FDMA and RRC pulse shaping still preserve their advantage compared to OFDM.

\begin{figure}
  \psfrag{-10*log10}[c][c]{\tiny $-10\log_{10} \sigma_\eta^2$}
\psfrag{BER}[c][c]{\tiny BER}
  \centerline{\includegraphics[width=5.5cm]{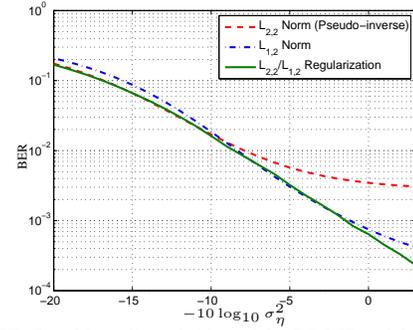}}
  \vspace{-0.5cm}
\caption{BER for $K = 5$ and $N = 30$, QPSK,  with 1-bit DACs.}
\label{fig_3}
\end{figure}

The BER performance is shown in Fig.~\ref{fig_2} and Fig.~\ref{fig_3} under constant envelope and 1-bit DAC constraints, respectively. In both cases, we observe substantial improvements at high SNR with sparse precoding. As explained earlier, the elastic net approach with the combined $\ell_1$ and $\ell_2$ regularization provides the best performance over the entire SNR range as it optimizes the trade-off between beamforming gain and PAPR reduction.   

\begin{figure}
  \psfrag{-10*log10}[c][c]{\tiny $-10\log_{10} \sigma_\eta^2$}
\psfrag{BER}[c][c]{\tiny BER}
  \centerline{\includegraphics[width=5.5cm]{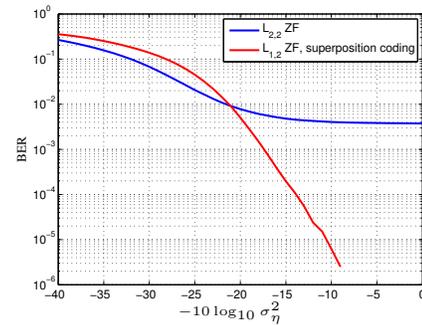}}
  \vspace{-0.5cm}
\caption{BER  for $K = 8$ and $N = 400$, 16QAM, 1-bit DACs}
\label{fig_5}
\end{figure}

Finally, we consider 16-QAM modulation with superposition coding and 1-bit DACs for $K=8$ and $N=400$ in Fig.\ref{fig_5}. At the receive side, we use the following blind estimation method for the scaling factor for each user prior to detection, as proposed in \cite{Hela2017}: 
\begin{equation}
f_k= T \cdot \frac{{\rm E}\left[|{\rm Re}\{s\}|+|{\rm Im}\{s\}|\right]}{ \sum_{t=1}^T  |{\rm Re}\{\hat{s}_k[t]\}|+|{\rm Im}\{\hat{s}_k[t]\}|},
\end{equation}
where $T$ is  the length the received sequence. Again, sparse linear precoding combined with superposition coding is clearly advantageous at high SNR. The proposed 16-QAM precoder design also better exploits the higher antenna count than standard ZF at higher SNR values.

\section{Conclusion}
\label{sec:copyright}
We introduced a novel sparse precoding method to cope with the PA and DAC issues in large-scale MIMO. We found the method to substantially improve the performance of standard linear precoders while reducing the amount of computation required to implement the linear precoder. With the increased demands for antenna element counts, we believe techniques offering reduced analog and digital hardware complexity such as the method presented here will be an essential ingredient in the development of future ultra-massive MIMO systems. 
 
\pagebreak
\bibliographystyle{IEEEbib}
\bibliography{IEEEabrv,references}

\end{document}